\documentclass[twocolumn,prd,nofootinbib,aps,prl,floats,floatfix,amsmath,amssymb,longbibliography,secnumarab
ic,superscriptaddress]{revtex4-2}
\usepackage[final]{graphicx}
\usepackage{hyperref}
\usepackage{amsmath}
\usepackage{bbm}
\usepackage{amsfonts}
\usepackage{amssymb}
\usepackage{latexsym}
\usepackage{graphicx}
\usepackage[english]{babel}
\usepackage{multirow}
\usepackage{float}
\usepackage{url}
\usepackage{slashed}
\usepackage{xcolor} 
\usepackage[utf8]{inputenc}
\usepackage{verbatim}
\usepackage{stackengine}
\usepackage[normalem]{ulem}

\newcommand{\be}{\begin{equation}}
\newcommand{\ee}{\end{equation}}
\newcommand{\ba}{\begin{array}}
\newcommand{\ea}{\end{array}}
\newcommand{\bea}{\begin{eqnarray}}
\newcommand{\eea}{\end{eqnarray}}
\newcommand{\sss}{\scriptscriptstyle}

\newcommand{\nn}{\nonumber}

\begin{document}
\title{Distortion of neutrino oscillations by dark photon dark matter}
\author{Gonzalo Alonso-\'Alvarez}
\email{galonso@physics.mcgill.ca}
\affiliation{McGill University, Department of Physics, 3600 University St.,
Montr\'eal, QC H3A2T8 Canada}

\author{Katarina Bleau}
\email{katarina.bleau@queensu.ca}
\affiliation{Queen's University,
    Department of Physics \& Engineering Physics Astronomy, Kingston, ON, K7L3N6, Canada}
\author{James M.\ Cline}
\email{jcline@physics.mcgill.ca}
\affiliation{McGill University, Department of Physics, 3600 University St., Montr\'eal, QC H3A2T8 Canada}

\begin{abstract}
A weakly coupled and light dark photon coupling to lepton charges $L_\mu-L_\tau$ is an intriguing dark matter candidate whose coherent oscillations alter the dispersion relations of leptons.
We study how this effect modifies the dynamics of neutrino flavor conversions, focusing on long baseline and solar oscillations.
We analyze data from the T2K, SNO, and Super-Kamiokande experiments in order to obtain world-leading limits on the dark photon gauge coupling for masses below $\sim 10^{-11}\,\mathrm{eV}$.
Degeneracies between shifts in the neutrino mass-squared differences and mixing angles and the new physics effect significantly relax the current constrains on the neutrino vacuum oscillation parameters.
\end{abstract}
\maketitle

\section{Introduction} Massive dark photons are a candidate for dark matter whose popularity has significantly grown in recent years.  
These particles may only interact with the visible sector through gravitational interactions, or through a small kinetic mixing with the SM photon.
However, the chances of discovering such particles are enhanced if they couple to a current involving standard model (SM) particles.
For theoretical consistency, such a current should be conserved, hence anomaly-free.
This limits the possible choices, and experimental searches very strongly constrain the couplings to first-generation quarks and leptons~\cite{Wise:2018rnb}. 
An attractive possibility is to couple to the difference between $\mu$ and $\tau$ lepton number, $L_\mu-L_\tau$, being anomaly-free and somewhat less stringently constrained~\cite{Foot:1990mn,He:1990pn,He:1991qd}. 
Furthermore this choice offers the possibility of
explaining the anomalous magnetic moment of the muon discrepancy in a simple way~\cite{Holst:2021lzm,Hapitas:2021ilr,Alonso-Alvarez:2021ktn}.

If the dark photon is sufficiently light, its cosmological population can be described by a classical field.
This classical field performs oscillations whose energy density behaves like pressureless matter\cite{Arias:2012az}.
For sufficiently small gauge coupling, the vector field is very weakly interacting and never thermalizes with the standard model bath.
The dark photon population may originate in the early universe from the vector misalignment mechanism~\cite{Nelson:2011sf,Arias:2012az,Alonso-Alvarez:2019ixv}, quantum fluctuations during inflation~\cite{Graham:2015rva,Ema:2019yrd,Alonso-Alvarez:2019ixv,Ahmed:2020fhc,Kolb:2020fwh}, or the decay of a scalar progenitor~\cite{Agrawal:2018vin,Co:2018lka,Bastero-Gil:2018uel,Long:2019lwl,Bastero-Gil:2021wsf}.
These production mechanisms predict different degrees of primordial polarization in the vector field, and it is unknown how such polarization evolves through cosmological structure formation.
For this work, we take the dark photon field to be at least locally polarized, so that we can take the field to have a fixed direction.

The coupling to the $L_\mu-L_\tau$ gauge boson leads to novel effects for $\mu$ or $\tau$ neutrinos.
The dispersion relation of these neutrino flavors is modified in the presence of the background dark photon field.
The leading effect can be described as a time-dependent effective neutrino mass, which affects their flavor oscillations. 
This was studied in~\cite{Alonso-Alvarez:2021pgy}, which focused on the possibility for resonant oscillations into a sterile flavor to occur in the early universe.
In that work, the impact of the dark photon coupling in the oscillations between active flavors was described and a preliminary study was made.
In this work, we extend this analysis to reinterpret data from long baseline and solar oscillation experiments.
For that, we numerically compute flavor survival probabilities in long-baseline experiments and in the sun, and compare the predictions with the measurements of the T2K, Super-Kamiokande, and SNO experiments.
Similar effects had been previously studied for light scalar particles $\phi(t)$ coupling to neutrinos~\cite{Berlin:2016woy,Brdar:2017kbt,Krnjaic:2017zlz,Liao:2018byh,Huang:2018cwo,Dev:2020kgz,Losada:2022uvr}, and vectors in the ultralight regime~\cite{Brdar:2017kbt}.
Recently, Ref.~\cite{Brzeminski:2022rkf} studied the vector case for the Super-Kamiokande and DUNE experiments, arriving at qualitatively similar results to us.
Quantitative differences are due to the distinct approaches and the fact that we consider the degeneracies between the vacuum oscillation parameters and the effect of the dark photon.

This paper is structured as follows. 
In section~\ref{sec:dp_nu_interaction}, we construct the modified Hamiltonian describing neutrino oscillations in the background dark photon field, and analytically solve the Schr\"odinger equation in the low- and high-frequency limits.
The following sections~\ref{sec:atmospheric} and~\ref{sec:solar} are devoted to the numerical solution of the Schr\"odinger equation for the case of long baseline and solar neutrino oscillations, respectively.
We use the results to place bounds on the dark photon and neutrino oscillation parameters using measurements from T2K, SNO, and Super-Kamiokande.
In section~\ref{sec:dp_parameter_space}, we discuss the impact of our study in the parameter space of $L_\mu-L_\tau$ dark photon dark matter.
We conclude in section~\ref{sec:conclusions}.

\section{Dark photon-neutrino interaction}\label{sec:dp_nu_interaction}
We are interested in a light massive gauge field $A'(t)$ that couples to
SM lepton doublets (and analogously for the corresponding right-handed leptons) via
\be
    {\cal L} = - g'A'_\mu\left[ \bar L_\mu\gamma^\mu L_\mu - \bar L_\tau\gamma^\mu L_\tau\right]
\ee
and makes up (a fraction of) the dark matter of the universe.
The cosmological evolution of a population of sufficiently light dark photons can be described as a classical field.
As such, considering $\vec A'$ as a background field that is varying slowly compared to the neutrino inverse energy, the dispersion relation of neutrinos gets shifted according to~\cite{Alonso-Alvarez:2021pgy}
\bea
    E_i &\to& \left( (\vec p \mp g'\vec A')^2 + m^2\right)^{1/2}\nn\\
        &\cong& |\vec p| \mp g' \hat p\cdot \vec A' + {m^2\over 2p} + O(g'^2A'^2)\,,
\eea
where the sign $\mp$ depends on the neutrino flavor, in the flavor basis.
The orientation of $\vec A'$ may vary on cosmological scales, but we will assume that it is constant in the vicinity of the solar system.  
Moreover the direction of the neutrinos in a given experiment is constant; either along the beam for long baseline experiments, or pointing from the sun for solar neutrino observations.  
The relative orientation of $A'$ and the neutrinos can vary with the rotation of the earth, around itself and around the sun. Hence $A'_\odot$ may also have a slow time dependence; however we will approximate it as being constant in the present study.
This is a good approximation for experiments that accumulate data in long time scales, as the effects of the modulations average out.
We thus define 
\begin{equation}
    \hat p\cdot \vec A' = A'_\odot\cos(m_{A'}t),
\end{equation}
where $A'_\odot$ denotes the amplitude of the dark photon oscillations in the solar neighborhood.
Using the standard prediction for the local dark matter density, $\rho_{\rm DM}^{\odot}\simeq 0.4\,\mathrm{GeV}/\mathrm{cm}^3$, and comparing with the average value $\rho_\mathrm{DM}^\mathrm{av}\simeq 1.3\,\mathrm{keV}/\mathrm{cm}^3$ from the Planck collaboration~\cite{Planck:2018vyg}, gives a DM concentration factor of $\rho_{\rm DM}^\odot / \rho_{\rm DM}^{\rm av}\sim 3\times 10^5$.
In terms of the contribution of the dark photon to the dark matter abundance of the Universe, we thus have
\begin{equation}\label{eq:A_sun}
    A'_\odot \simeq 25\,\mathrm{MeV}\left( \frac{10^{-10}\,\mathrm{eV}}{m_{A'}} \right) \sqrt{\frac{\Omega_{A'}}{\Omega_{\rm DM}}}.
\end{equation}

For realistic neutrino masses and mixings, it is necessary to envision some additional new physics to enable mixing between the different flavors, which is forbidden by the $L_\mu-L_\tau$ gauge symmetry.  A possible solution is to introduce two new Higgs doublets $H_\mu$ and $H_\tau$ that are appropriately charged under U(1)$_{L_\mu-L_\tau}$,
to allow the dimension-5 neutrino mass terms
\be
   \mathcal{L}\supset \frac12 \sum_{i,j = e,\mu,\tau}\Lambda_{ij}^{-1}(L_i H_i)(L_j H_j)
        +{\rm h.c.},
        \label{numass1}
\ee
where $H_e=H$ denotes the SM Higgs boson.  The vacuum expectation values (VEVs)  $v_\mu = \langle H_\mu\rangle$ and $v_\tau = \langle H_\tau\rangle$ contribute to the $A'$ mass
as $\delta m_{A'}^2 = g'^2(v_\mu^2 + v_\tau^2) \equiv g'^2 v'^2$. 
In section \ref{sec:dp_parameter_space} we discuss the interplay between the dark photon mass and coupling with the VEVs $v_{\mu,\tau}$ needed for the observed neutrino masses and mixings.
Other proposals to reconcile neutrino mixing and the $L_\mu-L_\tau$ gauge symmetry include extra Higgs doublets~\cite{Ma:2001md}, soft-breaking terms~\cite{Bell:2000vh,Choubey:2004hn}, and charged right-handed neutrinos~\cite{Binetruy:1996cs,Heeck:2011wj,Asai:2017ryy,Araki:2019rmw}.

For oscillations of relativistic neutrinos, we make the two-flavor approximation, considering either $\mu$-$\tau$ (atmospheric) or
$e$-$\mu$ (solar) oscillations.  The two-state system can be described 
by an unperturbed Hamiltonian $H_0 = E_0 + H_1$ where
\be
    H_1 = \left\{\begin{array}{ll} \frac{\Delta m^2_{23}}{4p}
    \begin{pmatrix}
    -\cos2\theta_{23} & \sin2\theta_{23}\\
    \sin2\theta_{23} & \cos2\theta_{23}
    \end{pmatrix},& \hbox{atmospheric}\!\!\!\!\!\!\!\!\!\!\\
   \frac{\Delta m^2_{12}}{4p} \begin{pmatrix}
    -\cos2\theta_{12} & \sin2\theta_{12}\\
    \sin2\theta_{12} & \cos2\theta_{12}
    \end{pmatrix},& \hbox{solar}
    \end{array}\right.
\ee  
The perturbation induced by the dark photon is given by
\be
    H_A = g'A_{\odot}'\cos(m_{A'}t)\left\{\begin{array}{ll}
        \begin{pmatrix}
    1 & \phantom{-}0\\
    0 & -1
    \end{pmatrix},& \hbox{atmospheric}\\
        \begin{pmatrix}
    0 & \phantom{-}0\\
    0 & -1
     \end{pmatrix},& \hbox{solar}
    \end{array}
     \right.
\label{NPcouplings}
\ee
To find the modified survival probability of a given neutrino flavor, it is necessary to numerically solve the Schr\"odinger equation
\be\label{eq:Schoredinger_eq}
    i{d\Psi\over dt} = H\Psi
\ee
for the two-state wave function $\Psi$.  For example, a neutrino that is initially in the flavor state $\nu_\mu$ has initial condition $\Psi(0) = (1,0)^{\sss T}$ in the $\mu$-$\tau$ basis.  The survival probability is given by
\be
    P_{\mu\to\mu}(t) = |\langle\Psi(0)|\Psi(t)\rangle|^2
\ee
from the numerical solution $\Psi(t)$.

\subsection{Analytic approximations}\label{sec:analytic_approximations}
The neutrino oscillation probability can be determined analytically in the limit of small or large $m_{A'}$.
If $m_{A'}\ll \Delta m^2/4p$, the perturbation can be treated adiabatically, giving rise to a slowly modulating
additional splitting between the diagonal mass elements in the vacuum oscillation probability.  We can parametrize the change using the dimensionless functions
\bea
    G &=& s_{ij} {4 p\over\Delta m_{ij}^2} g' A_\odot'\cos(m_{A'}t)\,,\nn\\
    F &=& \sqrt{1 -2G \cos 2\theta_{ij} + G^2},
    \label{eq:adiabatic_funs}
\eea
where $s_{23}=1$ for atmospheric oscillations and $s_{12}=1/2$ for solar ones, due to the different couplings in Eq.~\eqref{NPcouplings}.  Then one makes the replacements
$\Delta m_{ij}^2/4p \to F\,\Delta m_{ij}^2/4p$ and
$\sin 2\theta_{ij}\to \sin 2\theta_{ij}/F$ in the vacuum oscillation formulas.  For example, the survival probability becomes
\be\label{eq:adibatic_survival_probability}
    P_{\nu_i\to\nu_i}(L) = 1 - \sin^2 2\tilde\theta_{ij}
        \sin^2\left(L\,{\Delta m_{ij}^2\over 4p}\, F\right)
\ee
where
\begin{equation}
    \tilde\theta_{ij} = \frac{1}{2}\arcsin{\left(\frac{\sin{2\theta_{ij}}}{F}\right)}
\end{equation}
This expression is valid for oscillations over a distance $L$.  In this limit, there can be sensitivity to the phase of the DM oscillations, which we have absorbed into a shift of the time variable.

For large $m_{A'} \gg \Delta m^2/4p$, one can use second order perturbation theory on the Hamiltonian $H_A$, transformed to the eigenbasis of $H_1$.  Solving for the perturbations, one finds secularly growing terms, that can be resummed into a shift in the energies, which is equivalent to a reduction in the vacuum $\Delta m^2$, 
\begin{equation}
    \Delta m_{ij}^2 \to \Delta m_{ij}^2 \left(1 - \left(s_{ij} g'A'_\odot \sin 2\theta_{ij}\over 2 m_{A'}\right)^2\right).
    \label{eq:dm2shift}
\end{equation}
where $s_{ij}$ is as defined above.  This predicted behavior will be  validated below, in the numerical solutions to the full Schr\"odinger equation.

\begin{figure}[t]
\includegraphics[clip, width=\linewidth]{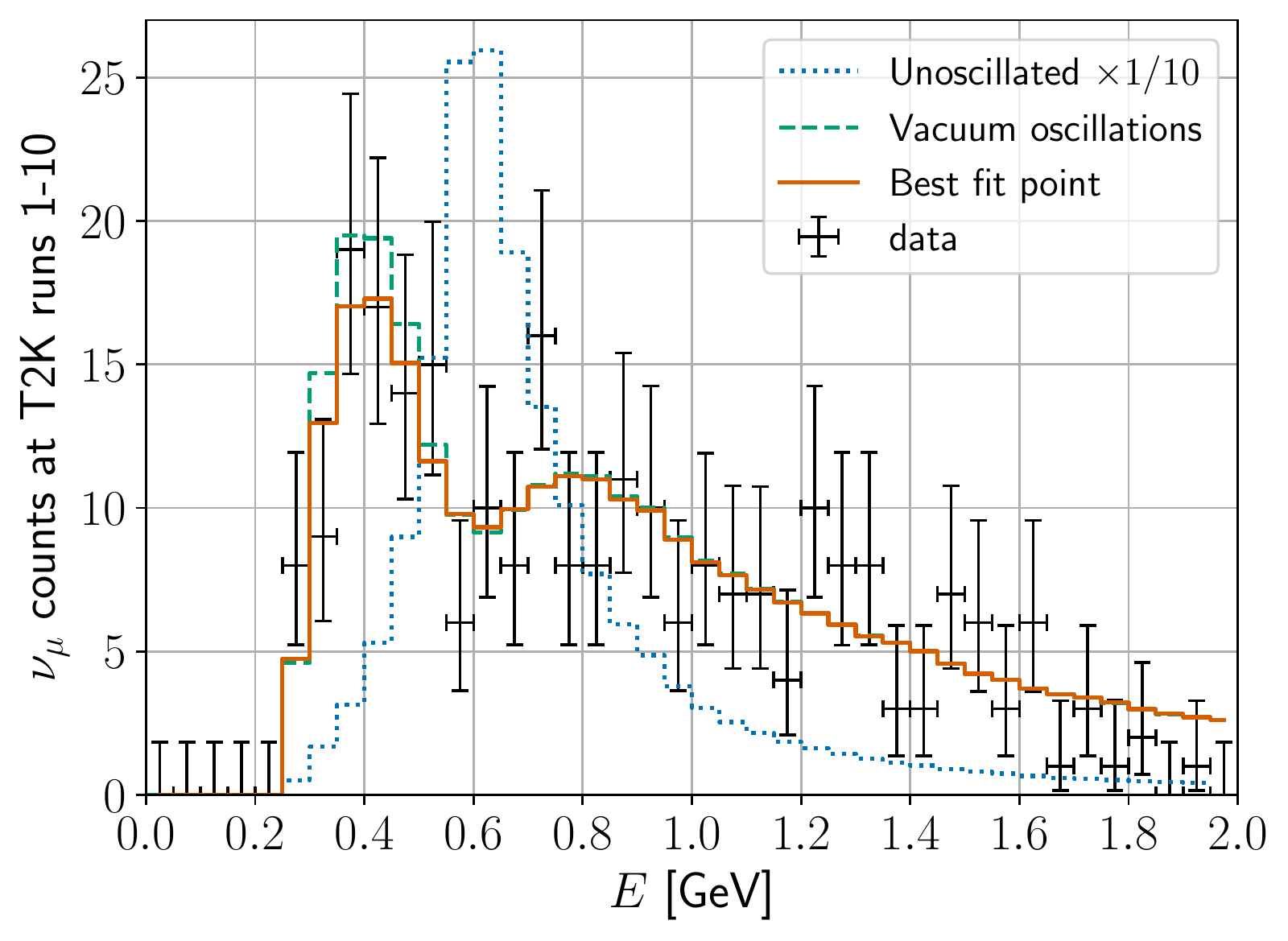}
\caption{Observed and predicted counts at T2K runs 1-10. The observed counts and their uncertainties are taken from Ref.\ \cite{T2K:2020run10}. The blue dotted line shows the expected signal in the absence of oscillations (rescaled by $1/10$), the green dashed line the standard oscillated spectrum in the absence of a dark photon ($\Delta m^2_{23}=2.5\times10^{-3}\,\mathrm{eV}^2$, $\sin^2\theta_{23}=0.55$), and the orange line the best fit with a dark photon ($\Delta m^2_{23}=3.4\times10^{-3}\,\mathrm{eV}^2$, $\sin^2\theta_{23}=0.45$, $m_{A'}=1.2\times 10^{-11}\,\mathrm{eV}$, $g'A'_\odot = 6.6\times10^{-21}\,\mathrm{GeV}^{-1}$).}
\label{fig:T2K_spectra}
\end{figure}

\section{Long baseline neutrino experiments}\label{sec:atmospheric}
The oscillations of $\nu_\mu$ into $\nu_\tau$ were originally observed within the atmosphere, using neutrinos generated by cosmic ray
interactions, and subsequently in long-baseline experiments.  
For our analysis, we use data from the the T2K (Tokai-to-Kamioka) experiment \cite{T2K:2011qtm}, which currently gives the best constraints on the atmospheric mixing parameters \cite{Esteban:2020cvm}.
T2K generates an initial beam of $\nu_\mu$ neutrinos and observes its flavor content at a detector $L= 295$\,km away.

For a fully quantitative understanding of T2K data,
one should consider oscillations of all three neutrino flavors.  However it is possible to describe the system in an effective two-flavor
formalism, which has also been employed by
T2K for some applications \cite{T2K:2021xwb}.
In this description, the dominant term in the $\nu_\mu$ survival probability is
\cite{Nunokawa:2005nx}
\be
    P_{\nu_\mu\to\nu_\mu}\cong
        1 - 4 |U_{\mu 3}|^2(1-|U_{\mu 3}|^2)
            \sin^2(L \Delta m^2_{\rm eff}/4p)
\label{eff2flav}
\ee
where $U_{\mu 3} = \cos\theta_{13}\sin\theta_{23}$
and $\Delta m^2_{\rm eff}$ is a weighted average of $\Delta m^2_{32}$ and $\Delta m^2_{31}$.  Since $\cos\theta_{13}\cong 0.99$, the prefactor
$4 |U_{\mu 3}|^2(1-|U_{\mu 3}|^2 \cong \sin^2 2\theta_{23}$.  Moreover $\Delta m^2_{\rm eff}$ is numerically close to $\Delta m^2_{23}$;
hence we use the vacuum mixing parameters $\theta_{23}$
and $\Delta m^2_{23}$ in the following two-flavor approximation, as a proxy for the more complicated
exact two-flavor expressions.  We will confirm that this
procedure provides a consistent description of the observed oscillations, and it allows us to use the analytic approximations derived in section \ref{sec:analytic_approximations} for limiting cases of the new physics parameters.  We defer a full three-flavor treatment to future work.

To compare to the experimental data, it is necessary to compute the survival probability $P_{\nu_\mu\to\nu_\mu}(E,t)$ at the detector position $t\cong L$, for a range of energies $E\lesssim 2\,$GeV, since the beam is not monochromatic.
The spectrum $dP/dE$ of the initial neutrino beam can be found in Ref.~\cite{T2K:2021xwb}.
For our calculations, we rebin this spectrum with the same energy bins as the measured event rate at the detector for observing runs 1-10, which was made public in Ref.~\cite{T2K:2020run10}.

\begin{figure*}[t]
\centering
\includegraphics[clip, width=\linewidth]{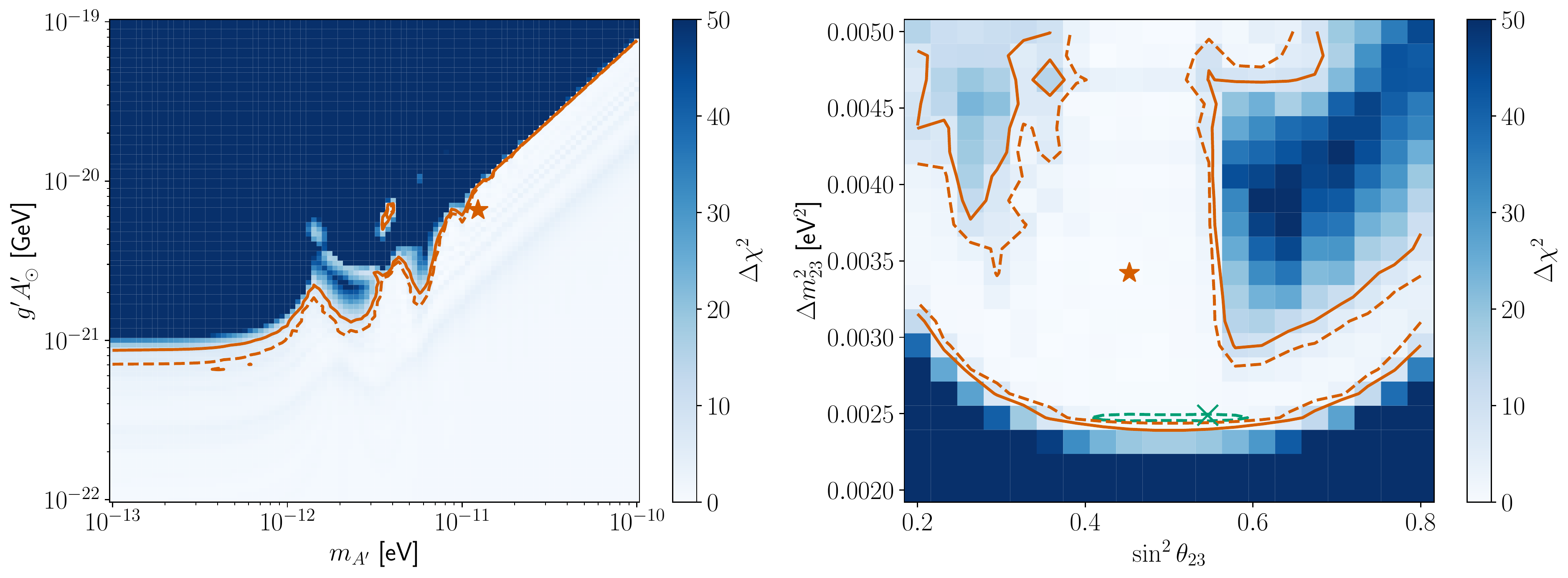}
\caption{Profiled likelihood functions from the T2K analysis. The solid (dashed) orange lines correspond to $95\%$ ($68\%$) c.l. limits, and the orange star represents the best fit point. For comparison, the green cross and the green dashed line show the best fit and $68\%$ c.l. contour, respectively, for the standard vacuum oscillation parameters in the absence of the dark photon.}
\label{fig:T2K_profiled_chi2}
\end{figure*}

To make the predictions for the final spectra in the presence of the dark photon  field, the first step is to calculate the muon neutrino survival probability.
For that, we numerically integrate the Schr\"odinger equation~\eqref{eq:Schoredinger_eq} to obtain the $\nu_\mu$ survival probability for each bin in the T2K energy spectrum.
The survival probability is then folded with the unoscillated spectrum to obtain the expected $\nu_\mu$ flux at the detector.
To compare with the observed count rate, we apply the detector response, including energy-dependent detection efficiencies and energy smearing effects.  
The detailed process is described in Appendix~\ref{sec:T2K_detector_response}.
The results for some exemplary spectra are shown in Fig.~\ref{fig:T2K_spectra}.

To study the impact of the dark photon in the oscillations, we perform a $\chi^2$ test comparing the data with the different predicted spectra.
We let four parameters vary in our fit: the vacuum oscillation parameters $\Delta m^2_{23}$ and $\sin^2\theta_{23}$, and the dark photon mass $m_{A'}$ and gauge coupling times field value in the solar neighborhood $g'A'_\odot$.
The resulting two-dimensional profile likelihoods in the dark photon and mixing parameters, respectively, are shown in Fig.~\ref{fig:T2K_profiled_chi2}.

The left panel of Fig.~\ref{fig:T2K_profiled_chi2} shows the constraints in the dark photon parameter space.
At low frequencies, $m_{A'}\ll \Delta m^2/4p \sim 10^{-12}\,\mathrm{eV}$, the survival probability is adiabatically modified as predicted in Eq.~\eqref{eq:adibatic_survival_probability}.
In this limit, the vacuum oscillation parameters are shifted by a quantity that depends on $g'A'_\odot$.
There is a slow modulation from the $\cos(m_A' t)$ in Eq.~\eqref{eq:adiabatic_funs} that may be relevant for observations on sufficiently long time scales.
Otherwise, the phase of the dark photon oscillation simply amounts to an $\mathcal{O}(1)$ multiplicative uncertainty in the limit on $g'A'_\odot$.
In our figures, we show the limits assuming $\cos(m_A' t)=1$, which maximizes the effect of the dark photon in the oscillations.
Thus, the constraint becomes asymptotically independent of $m_{A'}$.

At high frequencies, the effect of the dark photon is degenerate with a shift in $\delta m^2$, as shown in Eq.~\eqref{eq:dm2shift}.
The shift is proportional to $g'A'_\odot/m_{A'}$, which explains the slope of the constraint in Fig.~\ref{fig:T2K_profiled_chi2} for $m_{A'}\gg \Delta m^2/4p \sim 10^{-12}\,\mathrm{eV}$.
In the region of intermediate frequencies there is a nontrivial interplay between the neutrino and dark photon oscillation times, which results in the features seen in Fig.~\ref{fig:T2K_profiled_chi2} for $m_{A'}\sim 10^{-12}-10^{-11}\,\mathrm{eV}$.

The main effect of the dark photon on the neutrino vacuum oscillation parameters is to greatly enlarge the viable $\Delta m^2_{23}$-$\sin^2\theta_{23}$ parameter space.
In the right panel of Fig.~\ref{fig:T2K_profiled_chi2}, the orange lines correspond to the $1\sigma$ (dashed) and $2\sigma$ (solid) contours of the profiled likelihood, with the best fit point marked with an orange star.
For reference, the dashed green contour shows the $1\sigma$ region of the standard fit in the absence of the dark photon.
The improvement in the fit by the inclusion of the dark photon is not significant (a $\Delta\chi^2$ of $\sim 1$ at the best-fit point with $2$ extra parameters), but the likelihood becomes flatter along the $\Delta m^2_{23}$ and $\sin^2\theta_{23}$ directions.
This can be understood from the analytic approximations discussed in the previous paragraphs.
A larger value of $\Delta m^2_{23}$ can be compensated by the shift induced by the dark photon in the high-frequency limit.
Importantly, the inclusion of the dark photon never favors values of $\Delta m^2_{23}$ lower than the one corresponding to the standard vacuum fit. 
At low frequencies, both vacuum oscillation parameters are multiplicatively corrected, which allows for values of $\sin^2\theta_{23}$ far above and below the ones favored in the fit without the dark photon.

\section{Solar neutrino oscillations}\label{sec:solar}

Electron neutrinos with $\sim$MeV energies are copiously produced in the nuclear reactions occurring in the solar interior.
As they travel through the sun, the matter potential changes with the varying electron density, generating adiabatic conversions of $\nu_e$ into predominantly $\nu_\mu$ through the Mikheyev–Smirnov–Wolfenstein (MSW) effect~\cite{,Mikheyev:1985zog,Wolfenstein:1977ue}.
This produces a deficit in the expected solar $\nu_e$ flux at terrestrial experiments like the Sudbury Neutrino Observatory SNO~\cite{SNO:2002tuh} and Super-Kamiokande (SK)~\cite{Super-Kamiokande:2002ujc}.
This $\nu_e$ disappearance and its energy dependence allow to measure the solar oscillation parameters $\Delta m^2_{12}$ and $\sin^2\theta_{12}$.

\begin{figure}[t]
\centering
\includegraphics[clip, width=\linewidth]{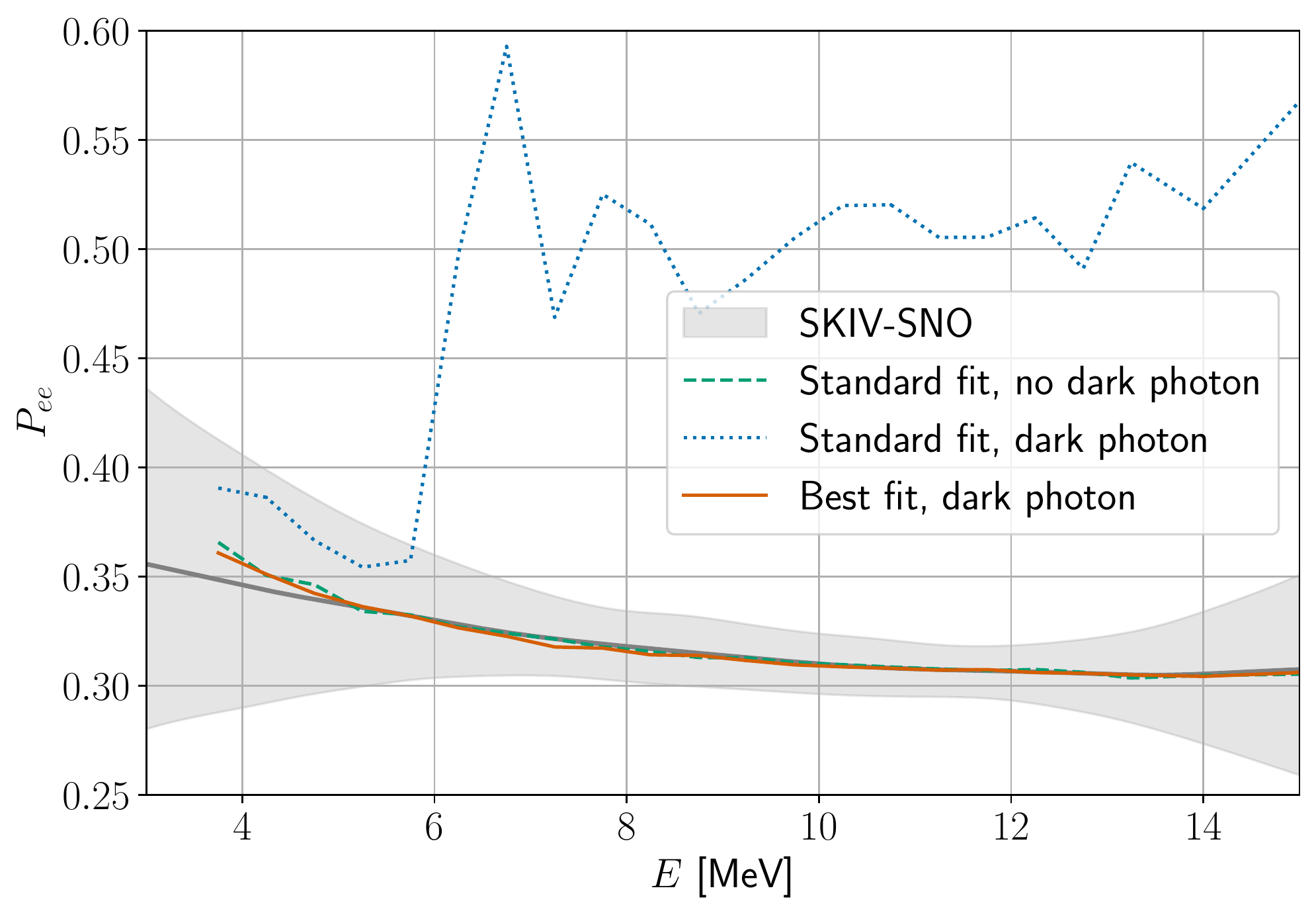}
\caption{Survival probability of $\nu_e$ emitted from the sun as a function of neutrino energy. The grey line and shaded region correspond to the best fit and $1$-$\sigma$ band from an analysis of the combined SNO~\cite{SNO:2011hxd} and SK-IV~\cite{Super-Kamiokande:2016yck} data.
The green dashed line shows the prediction of the standard two-flavor oscillations in the absence of a dark photon ($\Delta m^2_{12}=4.84\times10^{-5}\,\mathrm{eV}^2$, $\sin^2\theta_{12}=0.304$), while the dotted blue line shows the effect of a dark photon with $m_{A'}=2\times 10^{-12}\,\mathrm{eV}$ and $g'A'_\odot = 10^{-21}\,\mathrm{GeV}^{-1}$.
The orange line corresponds to the best fit with a dark photon ($\Delta m^2_{12}=6.0\times10^{-5}\,\mathrm{eV}^2$, $\sin^2\theta_{12}=0.303$, $m_{A'}=1.5\times 10^{-11}\,\mathrm{eV}$, $g'A'_\odot = 5.7\times10^{-21}\,\mathrm{GeV}^{-1}$).}
\label{fig:SKIVSNO_spectra}
\end{figure}

The presence of the oscillating dark photon background coupling to  $L_\mu-L_\tau$ modifies the $\nu_\mu$ matter potential and thus the above picture.
Our analysis of this effect is based on the legacy SNO results~\cite{SNO:2011hxd} and the SK-Phase IV ones~\cite{Super-Kamiokande:2016yck}.
The SK collaboration performed a combined fit to both datasets in Ref.\ \cite{Super-Kamiokande:2016yck} and provides constraints on the electron neutrino survival probability in the $3-15$~MeV energy window.
The result of this combined fit corresponds to the grey band in Fig.~\ref{fig:SKIVSNO_spectra}.

\begin{figure}[t]
\centering
\includegraphics[clip,width=\linewidth]{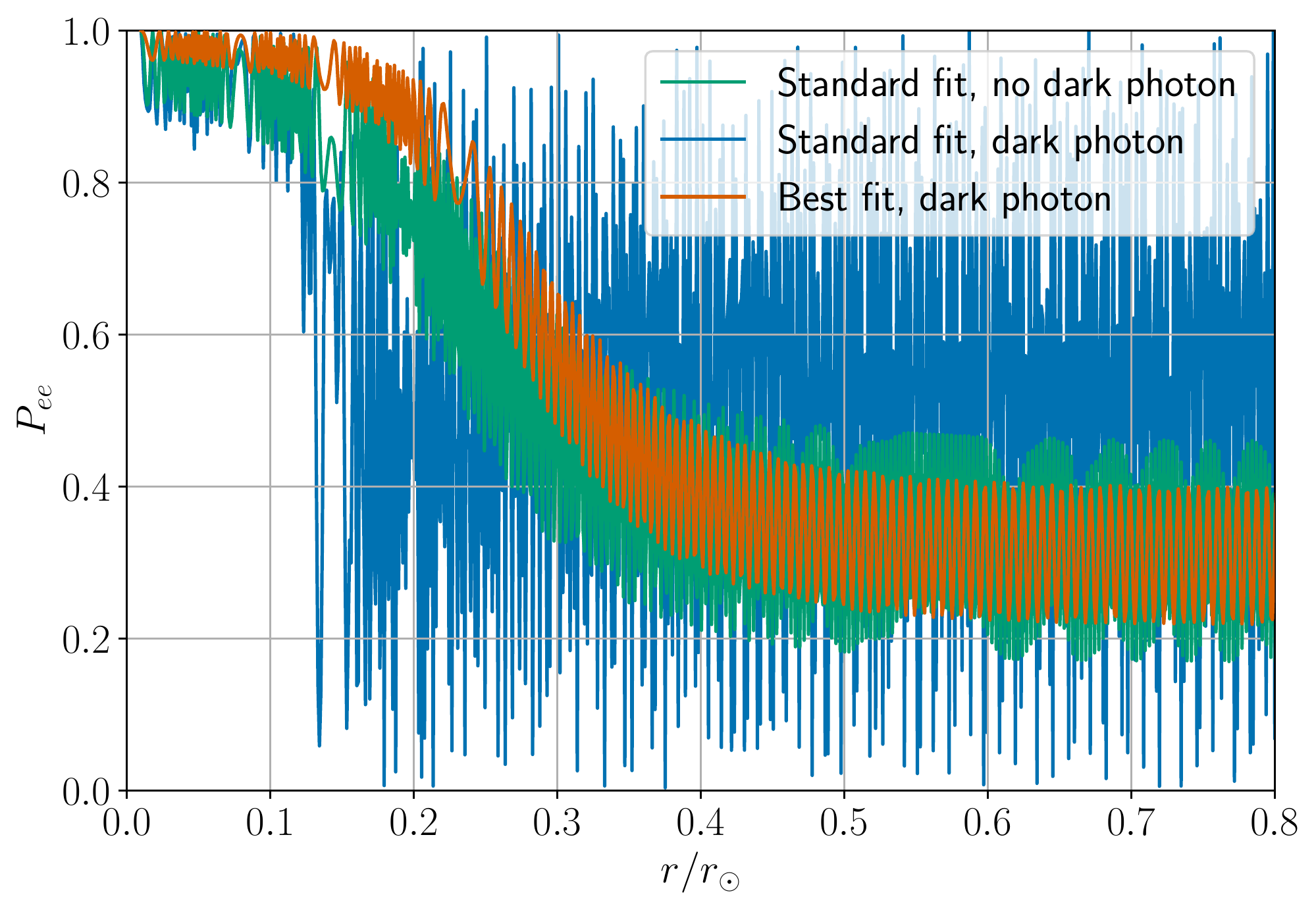}
\caption{
Electron neutrino survival probability along its trajectory from the center to the edge of the sun.
The three choices of parameters match those in Fig.~\ref{fig:SKIVSNO_spectra}. }
\label{fig:survival_probability_sun}
\end{figure}

To obtain our prediction for the survival probability, we numerically integrate the Schr\"odinger equation
\bea
    H &=& \frac{\Delta m^2_{12}}{4p}
    \begin{pmatrix}
    -\cos2\theta_{12} & \sin2\theta_{12}\\
    \sin2\theta_{12} & \cos2\theta_{12}
    \end{pmatrix}
    +
    \begin{pmatrix}
    \sqrt{2}G_F n_e & 0\\
    0 & 0
    \end{pmatrix}\nn \\
    &+& g'A_{\odot}'\cos(m_{A'}t)
    \begin{pmatrix}
    0 & 0\\
    0 & 1
    \end{pmatrix},
    \label{eq:solarH}
\eea
along the trajectory of a $\nu_e$ from the center to the outskirts of the sun.
Here, $G_F$ is the Fermi constant and $n_e$ is the electron density, which we calculate using the solar model given in Ref.\ \cite{Bahcall:2004pz}.
The survival probability $P_{ee}$ along the neutrino trajectory in the sun is shown in Fig.~\ref{fig:survival_probability_sun} for some exemplary parameter values.
In Fig.~\ref{fig:SKIVSNO_spectra}, we show the predictions for the observed survival probability as a function of energy for the same parameter choices.
These can be compared with the combined fit to the SNO SK-IV data.

\begin{figure*}[t]
\centering
\includegraphics[clip, width=\linewidth]{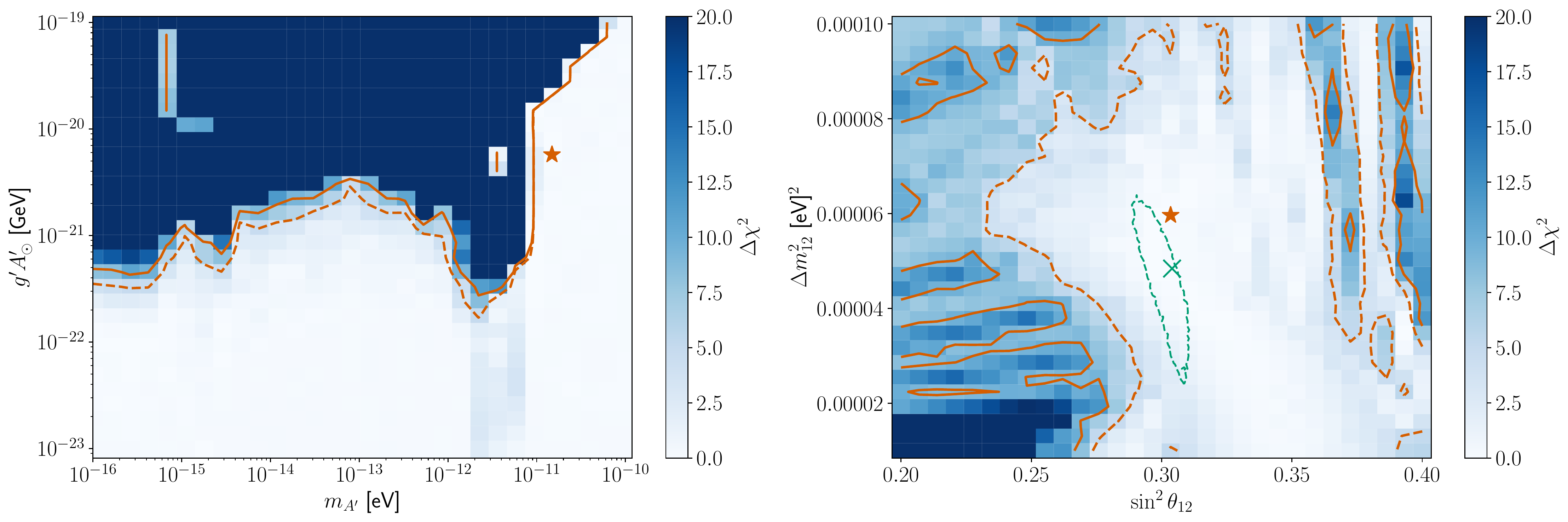}
\caption{Profiled likelihood functions of the combined SNO/SKIV fit. The orange solid (dashed) lines correspond to $95\%$ ($68\%$) c.l. limits, and the orange star represents the best fit point. For comparison, the green cross and the green dashed line show the best fit and $68\%$ c.l. contour, respectively, for the standard fit in the absence of the dark photon.}
\label{fig:SKSNO_profiled_likelihoods}
\end{figure*}

As in the long baseline case, we perform a $\chi^2$ fit to quantitatively compare our predictions with the measured survival probabilities, while varying
the vacuum oscillation parameters $\Delta m^2_{12}$ and $\sin^2\theta_{12}$, as well as the dark photon mass $m_{A'}$ and gauge coupling times field value in the solar neighborhood, $g'A'_\odot$.
The predicted $P_{ee}$ values are evaluated at the center of the SK-IV energy bins (see Table C.1 in~\cite{Super-Kamiokande:2016yck}).
The resulting profiled likelihoods in the two-dimensional parameter subspaces $(m_{A'},\,g'A'_\odot)$ and $(\Delta m^2_{12},\,\sin^2\theta_{12})$ are shown in Fig.~\ref{fig:SKSNO_profiled_likelihoods}.

The result is qualitatively similar to what was found for the atmospheric sector.
The main difference is that there are two characteristic frequencies involved in solar flavor transitions: the neutrino oscillation frequency given by $\Delta m^2/4p \sim 10^{-12}-10^{-11}\,\mathrm{eV}$, and the inverse time that the neutrino takes to traverse the sun, $1/R_\odot\sim 3\times 10^{-16}\,\mathrm{eV}$.
In the left panel of Fig.~\ref{fig:SKSNO_profiled_likelihoods}, a prominent feature in the profile likelihood is clearly visible when $m_{A'}$ matches the former range:
 the constraints become stronger by one to two orders of magnitude.
This is caused by a resonant behaviour of the neutrino flavor conversions when the neutrino oscillation frequency matches that of the dark photon field.

At low frequencies $m_{A'} \lesssim 10^{-12}\,\mathrm{eV}$, we recover the adiabatic limit discussed in Sec.~\ref{sec:analytic_approximations}.
Here the dark photon effect amounts to a multiplicative correction to the vacuum oscillation parameters.
This correction varies adiabatically with the instantaneous dark photon phase as the neutrino travels through the sun, and for each individual neutrino is sensitive to the initial dark photon phase.
For observations on sufficiently long time scales, the dependence on the initial phase averages out and only some mild features remain as seen for $10^{-16}\,\mathrm{eV} \lesssim m_{A'} \lesssim 10^{-12}\,\mathrm{eV}$ in the left panel of Fig.~\ref{fig:SKSNO_profiled_likelihoods}.
For masses below the inverse neutrino travel time in the sun, the constraint becomes asymptotically independent of $m_{A'}$ except for a slow modulation effect which may be relevant for observations on time scales shorter than $1/m_{A'}$.

At high frequencies $m_{A'} \gtrsim 10^{-11}\,\mathrm{eV}$, the dark photon oscillations average out and their leading effect is a shift in the vacuum oscillation frequency given by Eq.~(\ref{eq:dm2shift}).
This shift depends on the dark photon parameters through the ratio $g'A'_\odot / m_{A'}$, which causes the constraint to have the asymptotic slope seen in the left panel of Fig.~\ref{fig:SKSNO_profiled_likelihoods} in this regime.

The right panel of Fig.~\ref{fig:SKSNO_profiled_likelihoods} shows the impact of the coupling to the dark photon on the $\Delta m^2_{12}$-$\sin^2\theta_{12}$ parameter space.
For reference, we calculate the allowed $1$-$\sigma$ region in the absence of a coupling to the dark photon and show the result as a green dashed contour.
This $1$-$\sigma$ region faithfully reproduces the one found by the SK collaboration (see Fig.~34 in Ref.\ \cite{Super-Kamiokande:2016yck}).
The orange contours show the result of our fit in the presence of the dark photon.
The main consequence is that the viable region for the vacuum neutrino oscillation parameters are greatly enlarged, like in the long baseline oscillations case.
The small improvement in the likelihood for the best-fit point in the presence of the dark photon compared to the standard fit is not statistically significant.

\section{Dark photon parameter space}\label{sec:dp_parameter_space}
The constraints shown in Figs.~\ref{fig:T2K_profiled_chi2} and~\ref{fig:SKSNO_profiled_likelihoods} can be translated into limits on the combination of the $L_\mu-L_\tau$ gauge boson mass and coupling.
For that, we need to fix the value of the dark photon field $A'_\odot$ in the vicinity of the sun.
As discussed in Eq.~\eqref{eq:A_sun}, this value can be related to the fraction of dark matter that is made up by the dark photon.
Using Eq.~\ref{eq:A_sun} and assuming that the dark photon comprises the totality of the dark matter, we find the constraints shown in Fig.~\ref{fig:parameter_space_dark_photons}.
Our limits are the strongest for dark photon masses below $\sim 10^{-11}$~eV, except in the region already excluded by black hole superradiance.

A variety of previously existing astrophysical constraints on the gauge coupling $g'$ are shown in Fig.~\ref{fig:parameter_space_dark_photons}.
also taking into account the loop-induced kinetic mixing with $\epsilon=g^\prime/70$~\cite{Kamada:2015era}.
Bounds are shown from fifth forces between muons on neutron star binaries~\cite{KumarPoddar:2019ceq,Dror:2019uea} (NS binaries), modifications of the number of relativistic degrees of freedom during BBN~\cite{Huang:2017egl,Escudero:2019gzq,Dror:2020fbh} ($N_{\rm eff}>0.5$), observations of the duration of the neutrino burst from supernova SN1987A~\cite{Farzan:2002wx,Croon:2020lrf} (SN1987a), imprints of neutrino decay in the CMB angular and frequency spectra~\cite{Hannestad:2004qu,Hannestad:2005ex,Escudero:2019gfk,Escudero:2020ped,Barenboim:2020vrr} ($\nu$ decay CMB), and vector black hole superradiance~\cite{Baryakhtar:2017ngi,Cardoso:2017kgn,Cardoso:2018tly} (Superradiance).

The new constraints that we have derived from solar and atmospheric neutrino oscillations imply that the VEV contributing to the dark photon mass is bounded by
\be
    v'\gtrsim 15\,{\rm GeV}\,,
\ee
assuming that $m_{A'}$ comes from the Higgs mechanism.
Saturating this limit implies that the new Higgs doublets $H_\mu$ and $H_\tau$, introduced in Section
\ref{sec:dp_nu_interaction}, would get VEVs of the same order.  This can naturally accommodate the generation of charged $\tau$ and $\mu$ lepton masses,
as well as the neutrino masses via the seesaw mechanism, as indicated in Eq.\ (\ref{numass1}).

One implication of the above scenario is that $A'$ will mix at a small level with the standard model
$Z$ boson; but since the mixing is suppressed by $m_{A'}/m_Z \ll 1$, this does not have important phenomenological consequences, given that a minimum level of kinetic mixing is already expected.  Perhaps more importantly, the $A'$ remains massless at temperatures above electroweak symmetry breaking, which can impact its production mechanism.

Alternatively, the breaking of U(1)$_{L_\mu-L_\tau}$ can be accomplished by a complex scalar singlet $\phi$ with
$\langle\phi\rangle \sim 15\,{\rm GeV}$.  The required neutrino mass terms are obtained at the level of effective field theory by replacing $H_\mu \to \phi H/\Lambda$ and $H_\tau \to \phi^* H/\Lambda$ in Eq.\ (\ref{numass1}), for some scale $\Lambda$.  In this case no new Higgs doublets are required, and $m_{A'}$ can consistently arise at high scales
through the Higgs mechanism via $\langle\phi\rangle$.

\section{Conclusions}\label{sec:conclusions}

\begin{figure}[t]
\centering
\includegraphics[clip, width=\linewidth]{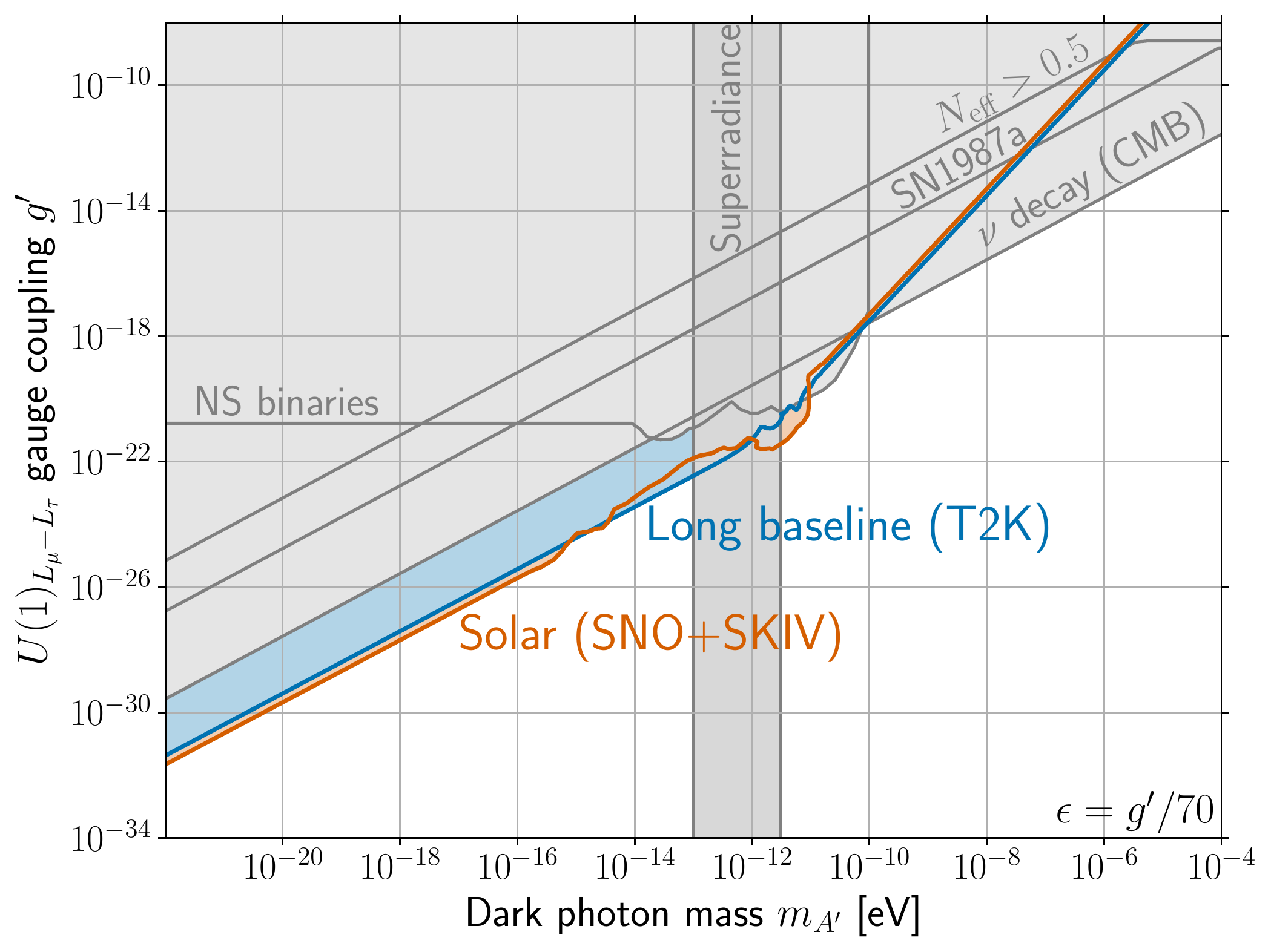}
\caption{Mass vs gauge coupling parameter space for $L_\mu-L_\tau$ dark photon dark matter.
The colored regions are the limits derived in this work and the grey ones correspond to other constraints. See the text for details.}
\label{fig:parameter_space_dark_photons}
\end{figure}

We have studied the impact of a background oscillating
$L_\mu-L_\tau$ dark matter gauge field on long baseline and solar neutrino oscillations. 
This yields constraints on the parameter space of the new physics, the mass $m_{A'}$ and gauge coupling $g'$ of the vector field, as shown in Fig.~\ref{fig:parameter_space_dark_photons}.
These constitute the strongest limits on the gauge coupling for most of the gauge boson masses below $\sim 10^{-11}$~eV.

Interestingly, the effect of the dark photon introduces novel degeneracies with the vacuum oscillation parameters of the standard model neutrinos.  
In the low frequency limit $m_{A'}\ll \Delta m^2/4p$, the dark photon multiplicatively corrects the vacuum $\Delta m^2$ and $\sin2\theta$.
In the high frequency limit $m_{A'}\gg \Delta m^2/4p$, the effect of fast $A'$ oscillations is indistinguishable from a reduction in the neutrino mass-squared difference.
For intermediate frequencies, the modifications are more intricate, as our numerical study for long baseline and solar oscillations shows.

A reanalysis of the data of the T2K, SNO, and Super-Kamiokande neutrino oscillation experiments including the background dark photon field has allowed us to delineate the range of parameters compatible with these experiments, which are shown in Figs.~\ref{fig:T2K_profiled_chi2} and~\ref{fig:SKSNO_profiled_likelihoods}.
Enlarging the allowed values of the vacuum oscillation parameters improves the consistency between the results of different neutrino oscillation experiments. For example, the KamLAND experiment~\cite{KamLAND:2013rgu} finds a best-fit value of $\Delta m^2_{12} = 7.53\times10^{-5}\,\mathrm{eV}^2$, which is at the edge of the 2-$\sigma$ region of the SNO/SK-IV fit. However, when the dark photon is included, both values become compatible at $1\,\sigma$.
As a next step, it would be interesting to include the dark photon in the analysis of the data from KamLAND and other reactor experiments.

The modification of neutrino flavor oscillations by the $L_\mu-L_\tau$ gauge field may have implications for other neutrino experiments beyond the ones studied in this work.
In particular, it may allow to ease existing tensions between different datasets, and potentially explain some of the standing experimental anomalies.
Furthermore, since the $\mu$ and $\tau$ neutrino masses receive an effective contribution that depends on the local dark matter density, their determinations by terrestrial experiments and cosmological observations may not necessarily match in the presence of the dark photon dark matter field.
We leave the study of these intriguing possibilities for future work.

{\bf Acknowledgments.}  We thank A.\ Konaka for helpful correspondence, and U.\ Rahaman for useful comments on the first version.  
This work was supported by NSERC (Natural Sciences and Engineering Research Council, Canada).
G.A. is supported by the Trottier Space Institute at McGill University through a McGill Trottier Chair Astrophysics Postdoctoral Fellowship.

\bibliographystyle{utphys}
\bibliography{ref}

\appendix
\section{T2K detector response}\label{sec:T2K_detector_response}

In order to obtain the expected signal at the T2K detector, we need to take into account detector efficiencies and energy resolution in addition to the neutrino survival probability.
We split this process into two steps.

\begin{figure}[t]
\centering
\includegraphics[width=\linewidth]{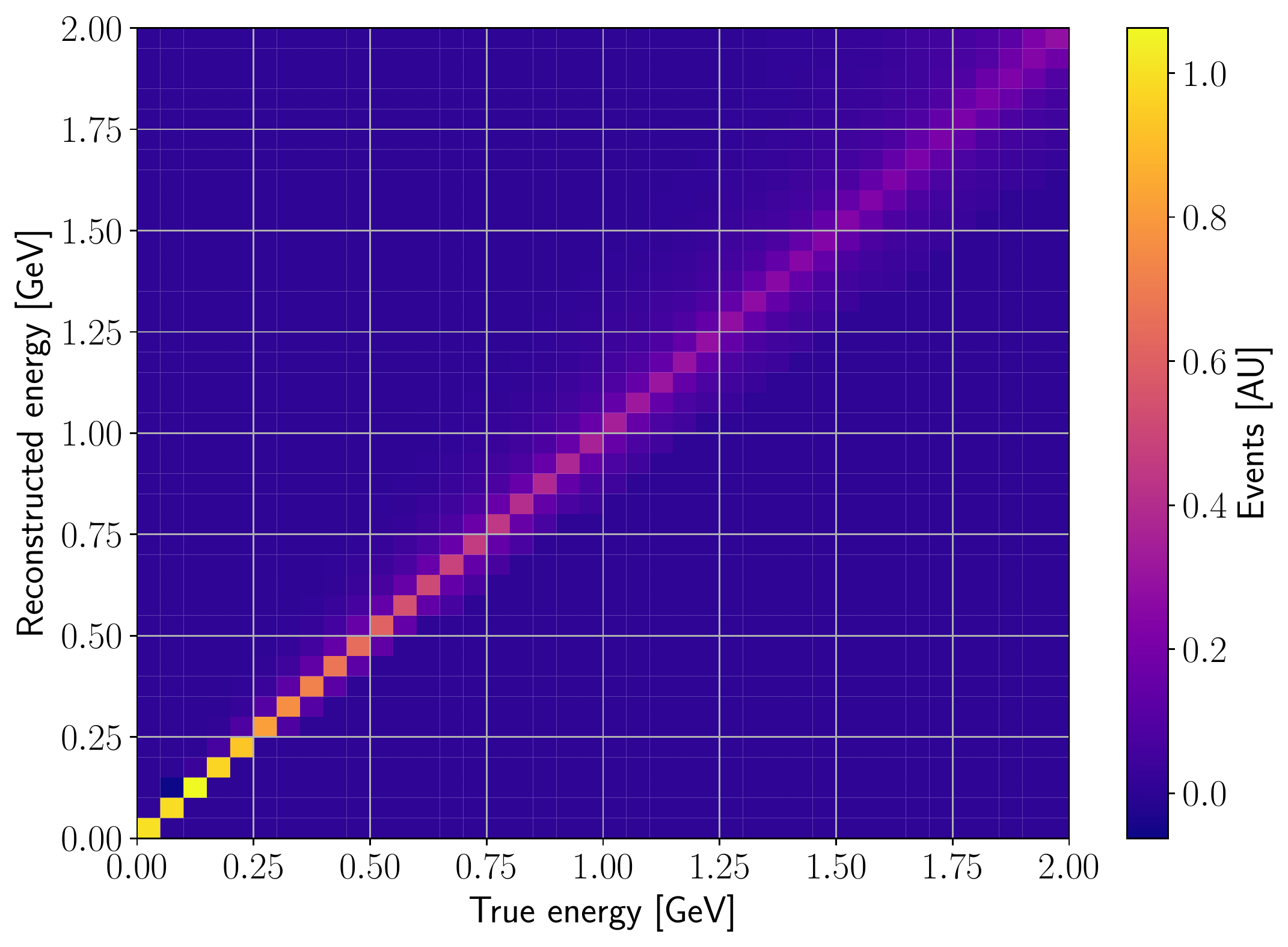}
\caption{Migration matrix showing the reconstructed energy bias at the Super-Kamiokande detector in the neutrino energy range relevant for the T2K experiment.
The effect of 1p1h and 2p2h interactions are taken into account.}
\label{fig:reconstructed_energy_bias}
\end{figure}

The first step is to construct the migration matrix that describes the smearing of the neutrino energy in the detector.
For that, we use the interaction model described in Ref.~\cite{T2K:2021xwb}.
The reconstructed energy bias at Super-Kamiokande is shown in Fig.~4 (bottom) of that reference for the two main interaction channels: one particle, one hole charged current interactions (1p1h), and two particles, two holes ones (2p2h). 
We neglect the effect of single pion production and deep inelastic scattering, which have much smaller cross sections in the energy range of interest.
From the reconstructed energy bias, we obtain the migration matrix shown in Fig.~\ref{fig:reconstructed_energy_bias}, which takes into account the interaction cross sections of the two channels.
The effect of smearing is significant for neutrinos with energies above $\sim 0.25\,\mathrm{GeV}$.

\begin{figure}[t]
\centering
\includegraphics[width=\linewidth]{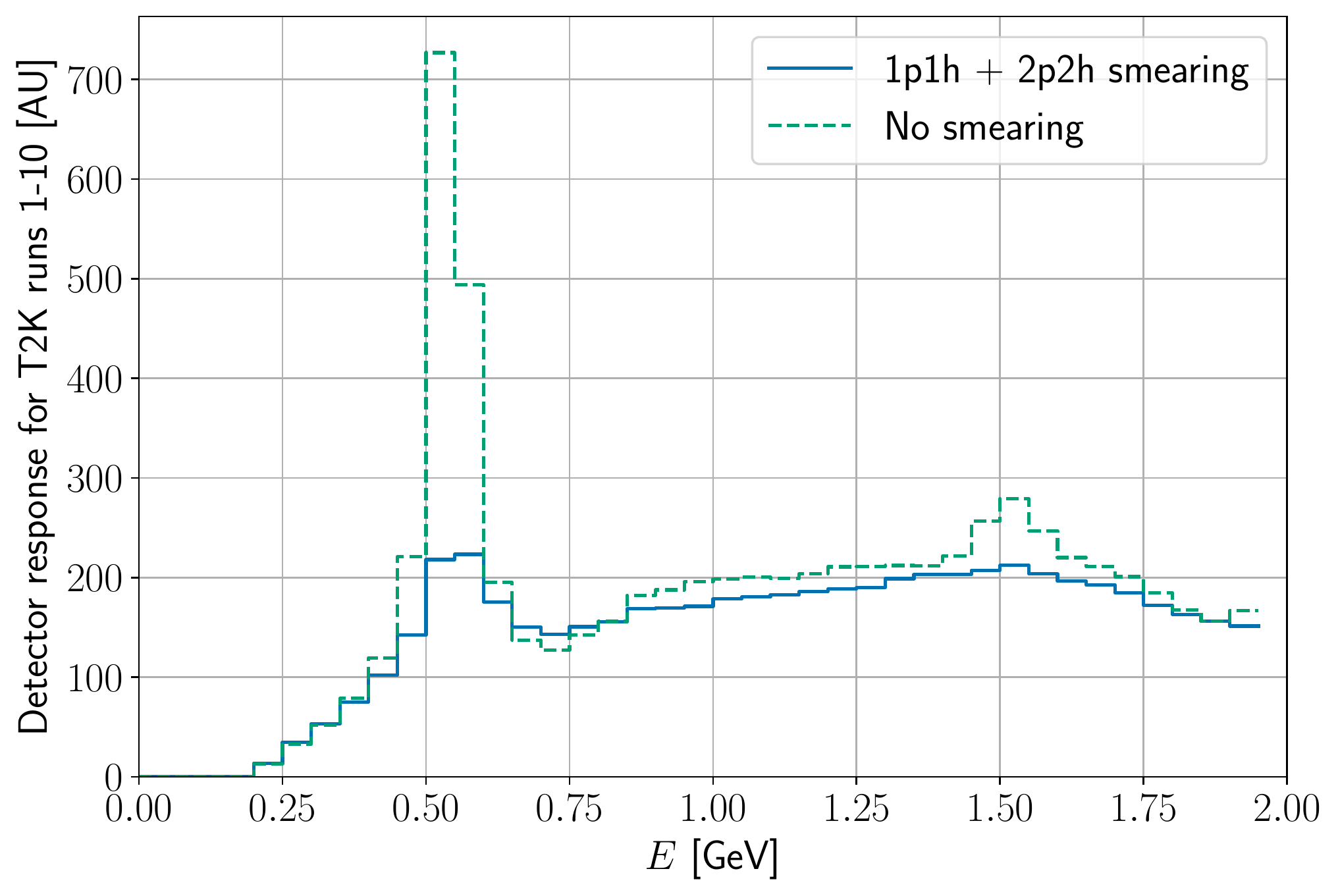}
\caption{Linear detector response of the Super-Kamiokande detector in the neutrino energy range relevant for the T2K experiment.
For reference, we show the detector response obtained without first applying the energy smearing to the simulated data.}
\label{fig:detector_response}
\end{figure}

The second step is to obtain the linear detector efficiency for each energy bin.
We infer this one using the expected signal published by the T2K collaboration in Ref.~\cite{T2K:2020run10}.
The detector response for each energy bin is obtained as the ratio of the expected signal by the oscillated neutrino flux.
We calculate the latter bin-by-bin by multiplying the initial flux given in Ref.~\cite{T2K:2021xwb} by the survival probability in the 2-flavor approximation with the T2K best-fit vacuum parameters $\Delta m^2_{23} = 2.49\times10^{-3}\,\mathrm{eV}^2$ and $\sin^2\theta_{23} = 0.546$.
Before taking the ratio, we perform a smearing of the energy using the previously calculated migration matrix.
This allows to take into account the reconstructed energy bias that cannot be captured by a linear per-bin detector response.
The result of this process corresponds to the blue curve in Fig.~\ref{fig:detector_response}.
To illustrate the importance of the energy smearing, we also show the dashed green curve corresponding to only applying a linear detector response.
In the absence of energy smearing, the detector efficiency is spuriously enhanced at the location of the first oscillation maximum, where two-flavor oscillations predict a reduction of the neutrino flux of $\sim 99\%$.
Correctly accounting for the energy smearing, events whose energy is misreconstructed populate the bins around the oscillation maximum and the flux is only reduced by $\sim 90\%$.

\newpage

\end{document}